\documentclass[twocolumn]{aastex631}
\begin{document}

% for float placement:
\renewcommand{\topfraction}{1.0}
\renewcommand{\bottomfraction}{1.0}
\renewcommand{\textfraction}{0.0}

\newcommand{\kms}{km~s$^{-1}$\,}
\newcommand{\msun}{$M_\odot$\,}
\newcommand{\masyr}{mas~yr$^{-1}$}

\shorttitle{The family of V1311 Ori}

\title{The family of V1311 Ori: a young sextuple system or a mini-cluster?}

\author{Andrei Tokovinin}
%\author[0000-0002-2084-0782]{Andrei Tokovinin}
\affiliation{Cerro Tololo Inter-American Observatory | NSF's NOIRLab
Casilla 603, La Serena, Chile}
\email{andrei.tokovinin@noirlab.edu}

\begin{abstract}
A compact  bound group of  four active M-type dwarfs  containing V1311
Ori is identified  in the Gaia catalog of nearby  stars.  Located at a
distance of  39 pc, it is  likely related to the  $\beta$ Pictoris and
32~Ori  moving  groups  by   kinematics,  isochronal  age,  and  other
indicators of youth (H$\alpha$ emission, presence of lithium, and fast
rotation).  The brightest star A is  a known close binary, for which a
preliminary 80-yr visual-spectroscopic orbit is determined.  Star B is
resolved here  into a 0\farcs08 pair,  and the faintest stars  C and D
are probably  single.  Considering the  non-hierarchical configuration
with projected  separations of  $\sim$10 kau, this  could be  either a
young sextuple system or a bound but dynamically unstable mini-cluster
(trapezium) that  avoided disruption  so far.  This  pre-main-sequence
system bridges the gap between moving groups and wide hierarchies.
\end{abstract}


%\maketitle

%---------------------------------------------------------
\section{Introduction}
\label{sec:intro}

Star formation  is a story  of concentration and dispersal,  of inward
(collapse) and outward  (jets and outflows) gas  motions.  Young stars
follow the pattern by condensing  into small groups and clusters which
later disperse, leaving behind bound stellar systems and single stars.
The  hierarchical collapse  can  last  for 10--30  Myr  at the  largest
spatial   scales,   but   it   is  much   faster   at   small   scales
\citep{Vasquez2019}.

Young moving groups (YMGs), such as $\beta$ Pictoris (BPMG), witness a
transition from  concentration to dispersal. Their  members still stay
together  in space  and preserve  coherent galactic  motion.  However,
smaller aggregates of stars, such as  wide pairs and multiples, may be
in the  process of disintegration  caused by their  internal dynamics,
and  in this  regard  are  similar to  dispersing  young clusters.   A
relatively frequent occurrence  of wide pairs (compared  to the field)
in   the    BPMG   and    in   other    YMGs   is    well   documented
\citep{Caballero2010,Alonso2015,Elliott2016}.  Although  the abundance
of wide pairs in the YMGs is uncontestable, their status is uncertain:
they could  be a mix of  long-lived bound binaries and  small unstable
disintegrating  groups  of  stars.   Ultra-wide pairs  are  even  more
frequent   in    the   1-Myr   old   Taurus    star-formation   region
\citep{Joncour2017};  they trace  the  primordial  clustering and  are
still in the concentration, rather than dispersal, phase.

\begin{figure}
\epsscale{1.2}
\plotone{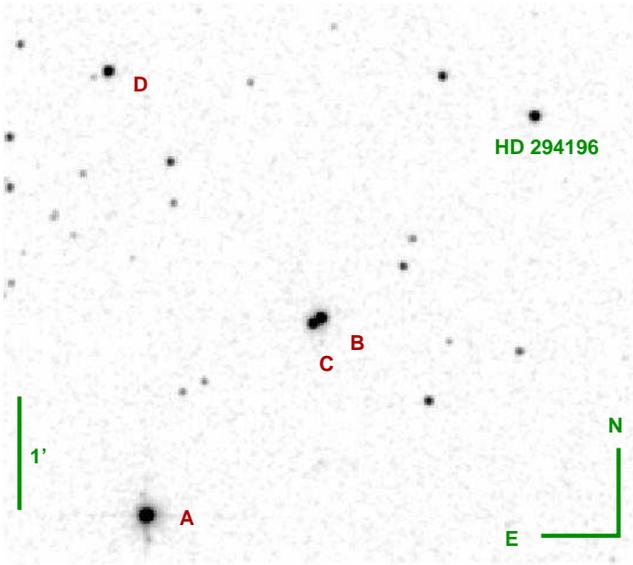}
%\plotone{Sky2.eps}
\caption{Location of the four stars  on the sky.  The underlying image
  is from 2MASS  band K \citep{2MASS}. The object is  2\degr ~north
  of the Orion nebula.
\label{fig:sky} 
}
\end{figure}

\begin{figure}
\epsscale{1.1}
\plotone{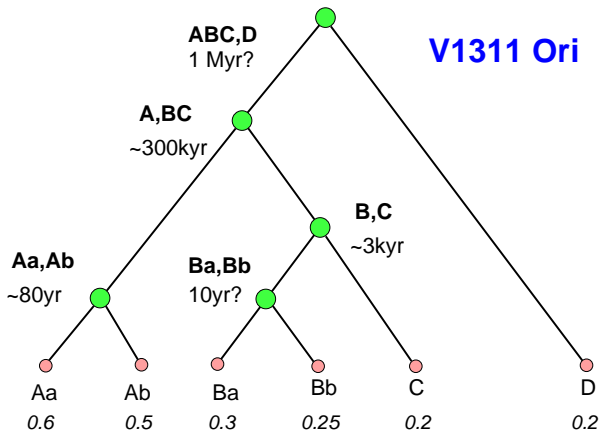}
%\plotone{mobile.eps}
\caption{Possible hierarchy of the V1311 Ori system. The numbers in
  italics are estimated masses of the components.
\label{fig:mobile}
}
\end{figure}

Here  I  study   a  group  of  four  co-moving  stars   in  the  solar
neighborhood, called V1311  Ori system after its  brightest member. It
has been  identified by search of  hierarchies in the Gaia  Catalog of
Nearby   Stars,  GCNS   \citep{GCNS}.    Some  of   these  young   and
chromospherically active stars were  studied individually as potential
members of YMGs and  for other reasons, but the fact  that they form a
gravitationally  bound  group has,  so  far,  escaped attention.   The
brightest stars  A and B  are close pairs,  so the system  contains at
least  six   components.   A  triangular  configuration   on  the  sky
(Figure~\ref{fig:sky}) with comparable separations  between A, BC, and
D  suggests   that  this  system  might   be  non-hierarchical,  hence
dynamically  unstable, which  makes it  particularly interesting.   As
shown  below,  currently  available   information  does  not  allow  a
conclusive choice  between the  two options, so  this system  could be
either   a   marginally   stable   hierarchy   or   a   disintegrating
mini-cluster. The first option is illustrated by the mobile diagram in
Figure~\ref{fig:mobile}.

Section~\ref{sec:par}  summarizes main  chracteristics of  these stars
and some published results relevant to  the nature and dynamics of the
group.   In  section~\ref{sec:speckle}  I report  new  high-resolution
imaging which  resolved the subsystem Ba,Bb  and allowed determination
of the  preliminary visual-spectroscopic  orbit of  Aa,Ab. Photometric
variability,  rotation,  and emission  lines  are  briefly covered  in
section~\ref{sec:ptm}.   Then  in  section\ref{sec:kin}  the  internal
motions  in this  system and  its relation  to YMGs  are investigated.
Discussion of the results in section~\ref{sec:sum} closes the paper.

%---------------------------------------------------------
\section{Main parameters and literature}
\label{sec:par}

\begin{deluxetable*}{l c  c  c c c} 
\tablecaption{Data on  components of the V1311 Ori system
\label{tab:1} }
\tablewidth{0pt}                                   
\tablehead{   
\colhead{Parameter} &
\colhead{A} &
\colhead{B} &
\colhead{C} &
\colhead{D} &
\colhead{E} 
}
\startdata
Simbad ID                & V1311 Ori   & PM 05319-0303W  & \ldots &    ESO-HA 737 &  RX J0534.0-0221 \\
2MASS                    & 05320450-0305291 & 05315786-0303367 &  05315816-0303397 & 05320596-0301159 & 05335981-0221325 \\
R.A. (EDR3)     & 05:32:04.51 &  05:31:57.88 & 05:31:58.17 & 05:32:05.97 & 05:33:59.83 \\
Dec. (EDR3)      & -03:05:30.0 & -03:03:37.6 & -03:03:40.7 & -03:01:16.8 & -02:21:33.3 \\
$\varpi$ (mas)\tablenotemark{a}           & 27.22$\pm$0.58\tablenotemark{b} & 26.30$\pm$0.09 & 25.91$\pm$0.02 & 25.96$\pm$0.02 & 29.11$\pm$0.03 \\
$\mu_\alpha^*$ (mas yr$^{-1}$) & 6.6$\pm$3.1\tablenotemark{b}      & 17.62 (4.2)      & 8.14      &       8.06 &  9.53  \\
$\mu_\delta$ (mas yr$^{-1}$)   & -51.1$\pm$3.2    & -51.79          & -54.02    &    -50.66  & -58.41    \\
RUWE                         &  33.50     & 3.95           & 1.32       & 1.23  &  1.21 \\
RV (km~s$^{-1}$)             & 22.2$\pm$0.3\tablenotemark{c}  & 23.1$\pm$1.0\tablenotemark{d}  & \ldots     & 23.6$\pm$2.7\tablenotemark{d} & 21.09$\pm$0.02\tablenotemark{e} \\
Spectral type               & M1.5V       & M4.5           & \ldots     & M5 & M3  \\
$V$ (mag)                   & 11.437      & 13.855         & 14.89      & 15.61 & 12.42 \\
$G$ (mag)                   & 10.442       & 12.701        & 13.241      & 13.912 & 11.267 \\
$J$ (mag)                   & 7.88         & 9.45          & 10.11      & 10.58 &  8.56 \\
$K_s$ (mag)                 & 7.01         & 8.54          & 9.22       & 9.70  & 7.70  \\
\enddata
\tablenotetext{a}{PMs and parallaxes  
  from Gaia EDR3 \citep{Gaia3}. }
\tablenotetext{b}{PM from Tycho-2 \citep{TYCHO}}
\tablenotetext{c}{RV of the center of mass Aa,Ab}
\tablenotetext{d}{RV from \citet{Bell2017}}
\tablenotetext{e}{RV from \citet{Fouque2018}}
%\tablenotetext{}{}
%\tablenotetext{}{}
%\tablenotetext{}{}
\end{deluxetable*}

The four  co-moving stars,  designated as A-D  in order  of decreasing
brightness, were found  in the GSNS. Their  mutual projected distances
are  within  the radius  of  10  kau  imposed  in the  initial  search
\citep{triads}.   To probe  for other,  more distant  members of  this
system, I  searched the  full Gaia Early  Data Release  (EDR3) catalog
\citep{Gaia3}  around  V1311  Ori  with  a radius  of  5\degr  ~and  a
constraint  on parallax  $\varpi >  20$\,mas. The  search returned  84
objects.  Filtering on  the proper motion (PM)  reduces this selection
to five stars. The additional fifth  star, called here component E, is
known  as  RX  J0534.0-0221  or  TIC  427346731,  and  it  is  another
well-studied member of BPMG.  The angular  distance between E and A is
52$'$ (0.6 pc), but E is closer to the Sun than ABCD by 4 pc (parallax
29.12$\pm$0.03 mas).  So, E is another member of the moving group, but
it is not bound to the V1311 Ori system.

Table~\ref{tab:1}  contains coordinates  and other  parameters of  the
components, including  star E for  completeness.  The first  two lines
give the Simbad and  2MASS \citep{2MASS} identifiers.  The coordinates
(equinox J2000, epoch 2016.0), parallaxes, and PMs are from Gaia EDR3.
The Reduced Unit  Weight Error (RUWE) parameter  indicates the quality
of Gaia  astrometric solutions,  being normally  below 1.4  for single
stars.  Elevated values  of RUWE for A  and B are caused  by motion in
the inner  subsystems Aa,Ab and Ba,Bb,  not accounted for in  the Gaia
5-parameter  astrometric  solutions.   This reduces  the  accuracy  of
parallaxes and biases  the PMs.  For this reason, the  long-term PM of
star  A  from  Tycho-2  \citep{TYCHO} is  preferred  over  the  biased
short-term PM  from Gaia; unfortunately,  no accurate long-term  PM is
available  for B.  The radial  velocities (RVs)  are  taken from  various
sources.

The  historical reasons  to study  these  stars were  their young  age
(manifested  by  X-ray  detections   and  H$\alpha$  emission),  their
potential  membership  in  YMGs,  or   their  proximity  to  the  Sun.
\citet{Finch2014} looked for stars within 25 pc from the Sun using the
UCAC4 astrometric catalog.  They list star D as UPM~0532-0303 and star
B as PM~05319-0303W and estimate their  distances as 20.7 and 16.3 pc,
respectively,  based on  photometric criteria  (these young  stars are
brighter   compared   to   normal  dwarfs,   causing   under-estimated
distances).  The  stars were subsequently  included in the  program of
astrometric     monitoring.      Its      results,     reported     by
\citet{Vrijmoet2020}, contain components  A, B, C under  a common name
UPM0531-0303 and with common coordinates, causing confusion.  In fact,
their components  A, B,  and C  correspond to  our stars  B, C,  and D
(Vrijmoet,  2021, private  communication).   The measured  parallaxes,
similar to the  Gaia ones, move these stars outside  the 25-pc horizon
of their program.  No astrometric perturbations were noted in the data
spanning 3.3 years.

The youth of V1311 Ori is  manifested by the H$\alpha$ emission in its
spectrum and by its X-ray detection.  However, originally the star was
attributed to the pre-main-sequence (PMS) population of the background
Orion  association, until  \citet{DaSilva2009}, \citet{Malo2013},  and
others considered V1311  Ori as a candidate member  of BPMG.  However,
\citet{Elliott2014} refuted  the BPMG  membership on  the basis  of RV
(biased  by  the  orbital  motion of  Aa,Ab).   \citet{Bell2017}  took
spectra of  stars B+C (blended) and  D, which they denoted  as THOR 33
and THOR 34,  respectively. They attributed these stars to  the 32 Ori
(THOR) moving group  which has age and kinematics similar  to the BPMG
and is  located at a  mean distance of  93\,pc, much further  than our
system.   They  detected  strong  H$\alpha$ emission  in  both  stars,
measured their RVs, and noted that D  had broad lines.  The two RVs of
D measured within a year agreed,  suggesting that it is a fast rotator
rather   than   a   close  binary.    \citet{Durkan2018}   took   five
high-resolution spectra  of V1311 Ori in  2011-2015, measured accurate
RVs, and detected an RV trend.   Their results are incorporated in the
orbital solution for Aa,Ab presented below.

\begin{figure}
\epsscale{1.1}
\plotone{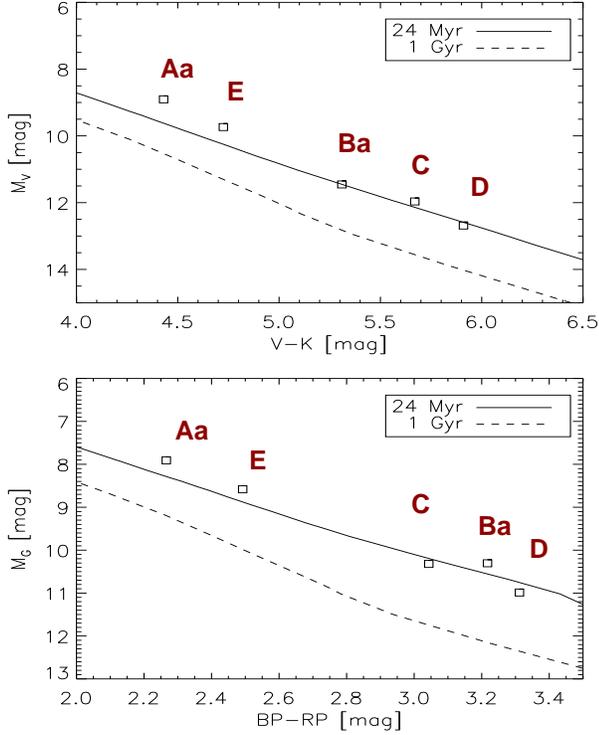}
%\plotone{CMD.eps}
\caption{Color-magnitude  diagrams.  The  full  and  dashed lines  are
  PARSEC isochrones \citep{PARSEC} for 24 Myr and 1 Gyr, respectively.
\label{fig:cmd}
}
\end{figure}

Figure~\ref{fig:cmd} shows the color-magnitude diagrams (CMDs) for the
members  of V1311  Ori family  and star  E in  the $M_V,V-K$  and Gaia
colors. The  contribution of companions  to the light  of A and  B has
been  subtracted.  The  stars  align  reasonably well  on  the 24  Myr
solar-metallicity isochrone  from \citet{PARSEC}, matching the  age of
BPMG and  THOR groups \citep{Bell2015,Bell2017}. The  masses estimated
from the  isochrone are  0.60, 0.26,  0.23, and  0.17 \msun  ~for Aa-D,
respectively.

%---------------------------------------------------------
\section{Inner Subsystems}
\label{sec:close}

%---------------------------------------------------------
\subsection{Speckle Interferometry}
\label{sec:speckle}

\begin{figure}
\epsscale{1.1}
\plotone{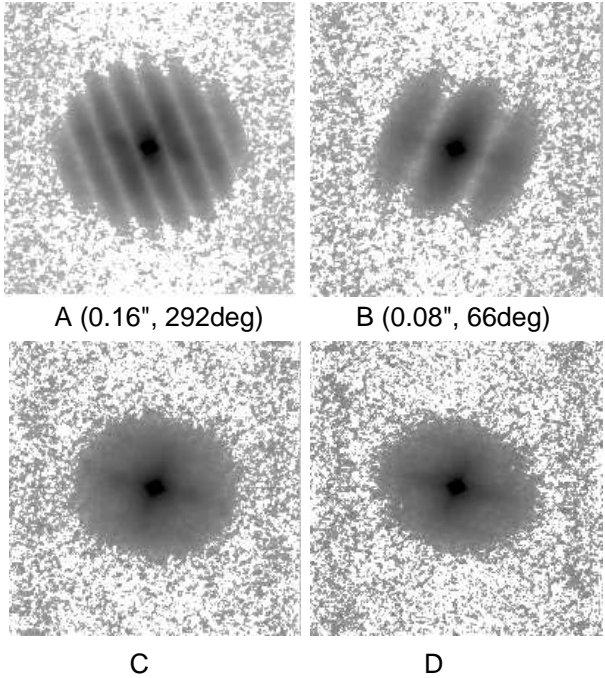}
%\plotone{Speckle2.eps}
\caption{Speckle  power  spectra  of   the  four  stars  (in  negative
  logarithmic  rendering)  recorded  on  2021 November 19  at  SOAR.
  Fringes in  the spectra of A  and B indicate that  they are resolved
  close pairs, while  C and D do not have  companions with separations
  above 0\farcs04.
\label{fig:speckle}
}
\end{figure}

The  components  of  this  system  were  observed  with  high  angular
resolution  at the  4.1  m Southern  Astrophysical Research  Telescope
(SOAR) in  2021 October-December.  The instrument  and data processing
are  covered in  \citep{HRCam}.  Briefly, series  of  400 images  with
exposure time  of 28 ms and  a pixel scale  of 15 mas are  recorded as
image  cubes  and processed  by  the  standard speckle  interferometry
method,  computing   the  spatial  power   spectrum,  auto-correlation
function,  and shift-and-add  image  (co-added with  centering on  the
brightest  pixel).   Two  data  cubes  per  observation  are  normally
recorded and  processed independently.   The stars were  observed here
with the  $I$ filter transmitting wavelengths  from 725 to 895  nm (at
half-maximum,   including  the   detector  response);   the  effective
wavelength  is  longer   than  824  nm  for  these   red  stars.   The
diffraction-limited resolution is about 40 mas.

Star A  was pointed and  resolved on  2021 October 18  (2021.80). This
pair is  listed in the  Washington Double Star Catalog  \citep{WDS} as
JNN~39. It has been resolved  for the first time by \citet{Janson2012}
and  later  confirmed  by  \citet{Janson2014}.  On  2021  November  19
(2021.89), I  observed all four  stars, taking advantage  of extremely
good 0\farcs5  seeing.  Star B  was also  resolved as a  tight binary,
while  stars C  and  D were  point sources.   Figure~\ref{fig:speckle}
shows  the speckle  power  spectra.   The faint  stars  B,  C, D  were
recorded  with a  50\,ms  exposure,  A ---  with  the standard  28\,ms
exposure.  The  pair Ba,Bb was  re-measured two days later  to confirm
its  resolution, and  both pairs  were measured  again on  December 16
(2021.96).  All SOAR speckle measurements of Aa,Ab and Ba,Bb are given
in Table~\ref{tab:speckle}.   The random errors of  relative positions
are 5 mas or less.  The quadrants of both pairs are determined without
the $180^\circ$ ambiguity.  The separation of Aa,Ab increases over two
months in agreement with the orbit presented below.

\begin{deluxetable}{c cccc} 
\tablecaption{Speckle interferometry of V1311 Ori
\label{tab:speckle} }
\tablewidth{0pt}                                   
\tablehead{   
\colhead{Pair} &
\colhead{Date} &
\colhead{$\theta$} &
\colhead{$\rho$} &
\colhead{$\Delta I $} \\
 & 
\colhead{(JY)} &
\colhead{ (deg) } &
\colhead{(arcsec)} &
\colhead{(mag)}  
}
\startdata
Aa,Ab & 2021.7983 & 291.3 & 0.1578 &   0.97 \\ 
Aa,Ab & 2021.8857 & 292.3 & 0.1655 &   0.87 \\ 
Aa,Ab & 2021.9596 & 292.9 & 0.1711 &   0.86 \\
Ba,Bb & 2021.8857 &  66.3 & 0.0816 &   0.52 \\ 
Ba,Bb & 2021.8910 &  65.5 & 0.0772 & 0.59   \\
Ba,Bb & 2021.9596 &  67.0 & 0.0824 & 0.55 
\enddata
\end{deluxetable}

%---------------------------------------------------------
\subsection{The Orbit of Aa,Ab}
\label{sec:orb}

\begin{figure}
\epsscale{1.0}
\plotone{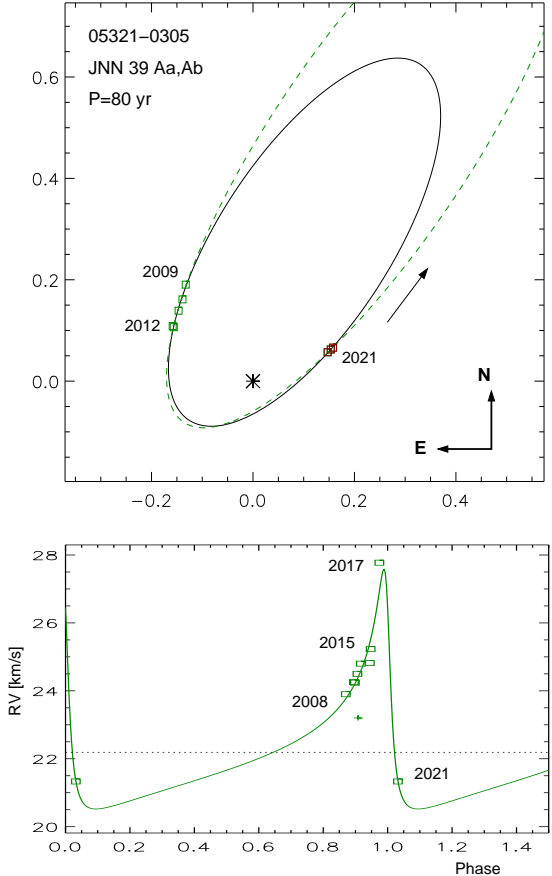}
%\plotone{Orbit.eps}
\caption{Tentative visual and spectroscopic orbit of Aa,Ab  (orbit
    1).  The  upper panel shows  position measurements with  scale in
  arcseconds (the SOAR measurements in  2021 are plotted in red, those
  of Janson et  al. in green).  The dashed line  shows the alternative
  long-period  orbit 2. The lower panel is  the RV curve (the dotted line
  shows  the   systemic  velocity,   the  cross  marks   the  rejected
  measurement).
\label{fig:orbit}
}
\end{figure}

%\input{vbtable.tex}
% visual
\begin{deluxetable*}{l cccc ccc cc}    
\tabletypesize{\scriptsize}     
\tablecaption{Two orbits of V1311 Ori Aa,Ab (JNN~39)
\label{tab:orb}          }
\tablewidth{0pt}                                   
\tablehead{
\colhead{Orbit} & 
\colhead{$P$} & 
\colhead{$T$} & 
\colhead{$e$} & 
\colhead{$a$} & 
\colhead{$\Omega_{\rm A}$ } & 
\colhead{$\omega_{\rm A}$ } & 
\colhead{$i$ }  &
\colhead{$K_1$ } & 
\colhead{$\gamma$ }  \\
&
\colhead{(yr)} &
\colhead{(yr)} & &
\colhead{(arcsec)} & 
\colhead{(deg)} & 
\colhead{(deg)} & 
\colhead{(deg)} &
\colhead{(km~s$^{-1}$)} &
\colhead{(km~s$^{-1}$)} 
}
\startdata
1 &  80  & 2019.35 & 0.778 & 0.499 & 133.2 & 47.6  & 63.0  & 3.52 & 22.20 \\
2  & 143 & 2019.56 & 0.854 & 0.807 & 132.5 & 45.6  & 67.8  & 3.88 & 21.76 
\enddata 
\end{deluxetable*}

The projected separation of Aa,Ab  corresponds to an orbital period on
the   order   of  20   yr.    Although   the  published   measurements
\citep{Janson2012,Janson2014}  and  the  SOAR  speckle  interferometry
cover only part of the orbit,  its general character is already clear.
The pair has passed through the  periastron in 2019 and now is opening
up again.

After fitting positional measurements by a set of prelimiary elements,
I included the  RVs as additional constraints and  fitted them jointly
with positions using  the IDL code {\tt ORBIT}  \citep{ORBIT}.  The RV
data   are  described   in  the   following  section.    The  position
measurements and RVs still do not  constrain the orbit well enough and
can be fitted by a family of orbits with periods ranging from $\sim$40
yr to over a century.  Two  representative orbits from this family are
listed in Table~\ref{tab:orb} in common notation.  The preferred 80-yr
  orbit  1 (Figure~\ref{fig:orbit})  is  obtained  by fixing  the
incination to  63\degr, to obtain the  expected mass sum of  1.1 \msun
for a parallax  of 26 mas.  The  second  orbit 2  with $P=143$ yr
and  a larger  eccentricity  results from  the  unconstrained fit  and
corresponds to the mass sum of  1.45 \msun, larger than estimated. The
fitted elements and their errors depend on the adopted data errors, (5
and 2 mas  for position measurements, 0.4 \kms for  RVs), so the large
formal  errors  of  the  elements are  esentially  meaningless.   Some
elements, e.g.   the periastron time  $T$ and the node  position angle
$\Omega_{\rm A}$, are already well  defined by the data.  The systemic
velocity  $\gamma$ of  22.20$\pm$0.28 \kms  derived for  the preferred
80-yr  orbit 1 is adopted as the RV of star A. The weighted rms residuals
to  both orbits are 1\,mas in positions and 0.3 \kms ~in RV,
  less than the estimated measurement errors.

The orbit  predicts that in  2016.0 Ab moved  relative to Aa  with the
speed of  $(-10.3, -33.4)$  \masyr ~in RA  and Dec,  respectively. The
difference between  the short-term PM of  A measured by Gaia  EDR3 and
the long-term PM in Tycho-2 is  $(3.5, 11.0)$ \masyr. The direction of
the  PM  difference   matches  the  expected  reflex   motion  of  the
photo-center and suggests that its amplitude is a factor of $f \approx
0.33$ smaller than the semimajor axis.  The estimated masses of Aa and
Ab  (0.6 and  0.5  \msun)  and the  magnitude  difference  of 0.9  mag
measured  at SOAR  correspond to  $f=0.15$. This  factor increases  to
$f=0.22$   if  a   larger   $\Delta   m  =   1.3$   mag  measured   by
\citet{Janson2012} is adopted.

The orbital inclination,  RV amplitude, and the mass  of Aa correspond
to the mass of 0.38 \msun for Ab, somewhat smaller than estimated from
the absolute magnitude and the isochrone.  Considering the preliminary
nature  of the  Aa,Ab orbit,  it is  premature to  investigate further
these  minor   disagreements  between  the  photo-center   motion,  RV
amplitude, and estimated masses.  Although  the distance to the system
is  known  quite  well,  the  orbit is  not  yet  useful  for  testing
evolutionary models of low-mass PMS stars.

%---------------------------------------------------------
\subsection{Photometry and Spectroscopy}
\label{sec:ptm}

\begin{figure}
\epsscale{1.1}
\plotone{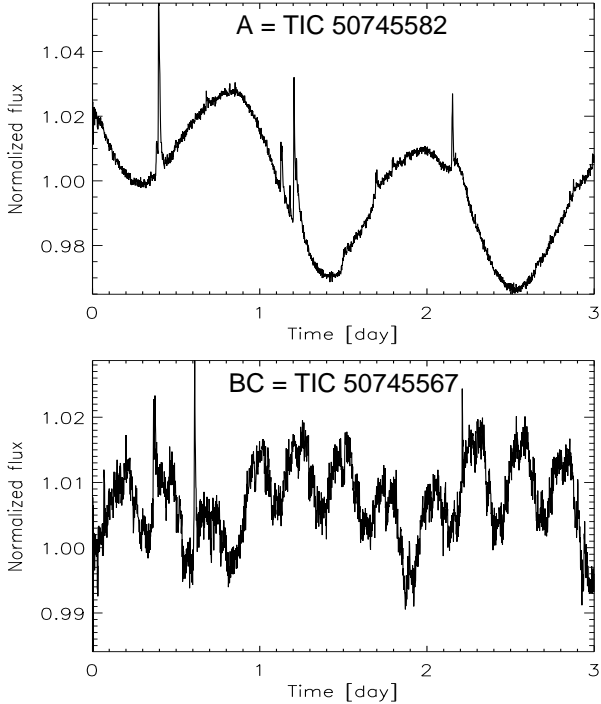}
%\plotone{Tess.eps}
\caption{Fragments of the  light curves of stars A and  BC recorded by
  TESS  in  sector 32  (2020  November).  The  first  point is  at  JD
  2459174.2234.
\label{fig:TESS}
}
\end{figure}

Stars A and  B are present in  the TESS input catalog  as TIC 50745582
and 50745567, respectively.   Their fluxes were monitored  by the TESS
satellite  \citep{TESS} in  sectors  6 (2018  November)  and 32  (2020
November).     I    extracted    the    light    curves    from    the
\href{https://mast.stsci.edu/}{MAST}  archive;  their 3-day  fragments
are plotted in Figure~\ref{fig:TESS}.  Star  A shows an almost perfect
sinusoidal variation with  a period of 1.119 days and  an amplitude of
0.020 (a  weak second harmonic is  detectable in the 2020  data), with
frequent  flares. There  is  a  second period  of  4.37  days with  an
amplitude of  0.015, previously detected from  ground-based photometry
by \citet{Messina2017}, so A is a multi-periodic M-dwarf as defined by
\citet{Rebull2018}.  The flux variation of star B (blended with C) has
the main  period of 0.2642 days  (6.34 h), implying rotation  near the
breakup speed, similar to some  young low-mass stars studied by Rebull
et al.   The light  curve is  not sinusoidal,  resembling scallop-type
variable late-M  dwarfs identified by \citet{Stauffer2017}.   The flux
of BC recorded by TESS corresponds to  at least three stars Ba, Bb, and
C.  The period of 1.119 days is also detectable in the flux of BC with
an amplitude  of 0.005,  presumably due to  contamination from  star A
located at 148\arcsec ~from BC.

\begin{figure}
\epsscale{1.1}
\plotone{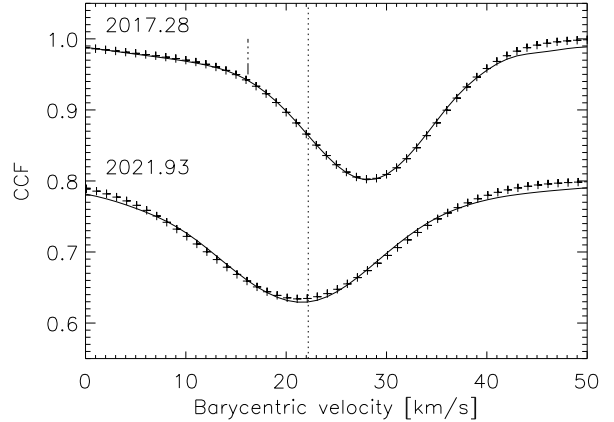}
%\plotone{ccfplot.ps}
\caption{CCFs  of the  FEROS and  CHIRON  spectra of  V1311 Ori  (full
  lines) and  their Gaussian approximations (crosses).  The CHIRON CCF
  is displaced vertically  by $-$0.2.  The vertical  dotted line marks
  the systemic velocity of Aa,Ab,  the short dash-dot line corresponds
  to the expected RV of the secondary component in 2017.28.
\label{fig:CCF}
}
\end{figure}

In the  orbital fit,  I used five  RVs measured  by \citet{Durkan2018}
from high-resolution spectra taken with FEROS (Fiberfed Extended Range
Optical Spectrograph)  on the ESO-MPG  2.2 m telescope from  2010.9 to
2015.098. I also found in the ESO archive another FEROS spectrum taken
on JD  2457855.5019 (2017.279) and  measured the  RV of 27.77  \kms ~by
cross-correlation. Three RVs from table C.3 of \citet{Elliott2014} are
also used  (the first one in  2008.9), although one discrepant  RV (JD
2455904.18) is  excluded from  the fit  (cross in  the lower  panel of
Figure~\ref{fig:orbit}). I  re-computed the RV from  that spectrum and
obtaied  a  similar  result,  suggesting a  problem  with  wavelength
calibration. 

A contemporary spectrum of V1311 Ori  was taken on 2021 December 6 (JD
2459555.7431) using the CHIRON high-resolution optical spectrometer on
the Cerro Tololo  1.5 m telescope, operated by  the SMARTS consortium.
The    instrument   and    data    processing    are   described    in
\citep{CHIRON,Paredes2021}.  The  spectrum was  acquired in  the fiber
mode with a resolution of 27,000 and an exposure time of 15 min.  The
RV of  21.33 \kms  was determined  by computing  the cross-correlation
function  (CCF)  with  a  binary  mask based  on  the  solar  spectrum
\citep[see  details in][]{Tokovinin2016}.   Figure~\ref{fig:CCF} shows
the CCFs for  the last FEROS spectrum taken in  2017.27 (near the peak
of the RV curve)  and the CHIRON spectrum.  Both have  one dip with an
rms width of 7.49 and 9.13 \kms, respectively (the CHIRON spectrum has
a  lower resolution  compared  to  FEROS).  The  dip  width implies  a
projected rotation velocity $V \sin i$ of 12 \kms; \citet{DaSilva2009}
also measured a rotation velocity of 12 \kms. 

A  star like  Aa  with  a 0.77  $R_\odot$  radius  (inferred from  the
isochrone) rotating at  a 1.119 day period has  an equatorial velocity
of 34  \kms. The CCF dip  indicates a rotation three  times slower and
matches the longer photometric period  of 4.37 days. Most likely, star
Ab is a fast rotator responsible  for the 1.119-day period.  Its broad
and low-contrast dip  is not detected in the 2017  CCF at the expected
velocity  of  16.2  \kms  (the short  line  in  Figure~\ref{fig:CCF}),
despite the moderate  magnitude difference between Aa  and Ab measured
by  speckle interferometry.   However,  this CCF  does  have a  slight
asymmetry and  it was tentatively  fitted by two  Gaussians (crosses),
with  Ab rotating  at  $\sim$20  \kms and  having  a  dip contrast  of
0.03. Unfortunately, the dips of Aa and Ab are heavily blended now and
will remain blended for decades, until the next periastron.

Spectrum  of star  B was  taken with  CHIRON on  2021 December  22 (JD
2459571.6361). Its CCF has a shallow and wide dip corresponding to the
RV of 17.5$\pm$1 \kms  and an rms width of 28 \kms, or  $V \sin i \sim
55$ \kms.  I  also correlated this spectrum with  synthetic spectra of
late-M dwarfs  to confirm these measurements.   The photometric period
of 0.2462  days detected in BC  corresponds to an equatorial  speed of
112 \kms (for  a 0.55 $R_\odot$ radius). Low-resolution  spectra of BC
and D  taken by \citet{Bell2017}  show very  broad lines in  D, rather
than in BC,  but the CHIRON spectrum of B  confirms its fast rotation;
the lines  of Ba and Bb  are totally blended. Orbital  motion of Ba,Bb
should cause the RV variation.

\begin{figure}
\epsscale{1.1}
\plotone{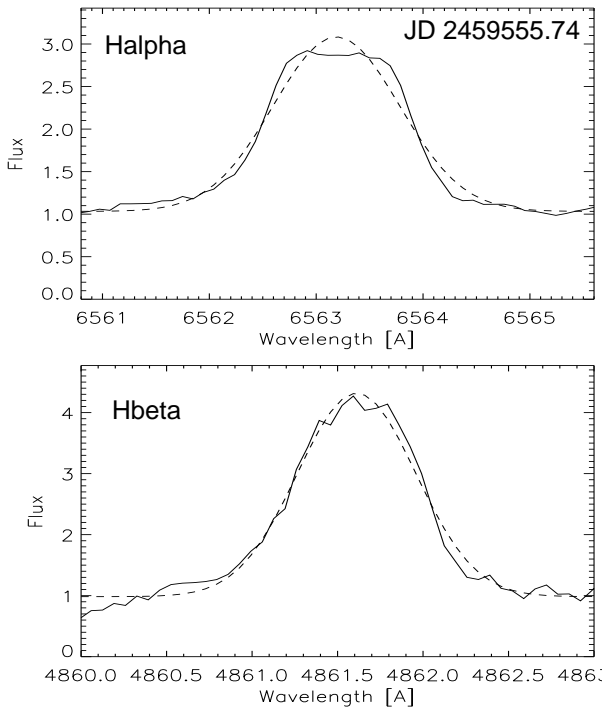}
%\plotone{Bahlmer.eps}
\caption{Hydrogen emission  lines in the CHIRON  spectrum (full lines)
  and fitted Gaussians (dashed lines).
\label{fig:Ha}
}
\end{figure}

Several authors noted other signs of  youth in the spectra of stars A,
B, and D, namely presence of the lithium line and a strong emission in
the Balmer hydrogen lines.   Figure~\ref{fig:Ha} shows these emissions
in the CHIRON  spectrum of A.  The equivalent widths  of H$\alpha$ and
H$\beta$ are  $-$3.13 and $-$2.78  \AA, respectively, their  FWHMs are
1.39 and 1.86  \AA ~(85 \kms).  The CHIRON spectrum  of V1311 Ori also
shows chromospheric  emission in the cores  of the sodium D  lines. In
the  FEROS spectra,  the  equivalent width  of  H$\alpha$ varies  from
$-3.72$  to  $-2.75$  \AA,  and its    double-peaked  profile  is
slightly  asymmetric, with  the maximum  on  the left  side. A  strong
H$\alpha$ emission is present in the CHIRON spectrum of B.

%---------------------------------------------------------
\section{Hierarchy or Cluster?}
\label{sec:kin}

\begin{figure}
\epsscale{1.1}
\plotone{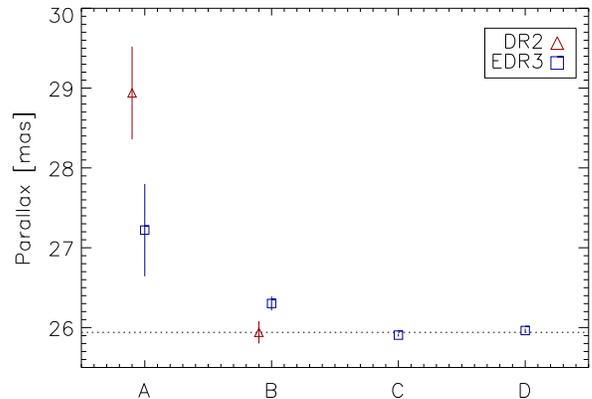}
%\plotone{plxplot.ps}
\caption{Parallaxes of four stars in Gaia EDR3 and DR2. The dotted
  line marks the mean parallax of C and D, 25.94\,mas. 
\label{fig:plx}
}
\end{figure}

In this  Section, two alternative  views of  the V1311 Ori  family are
presented. The choice depends on the reliability of Gaia parallaxes of
stars A  and B.   The EDR3  parallaxes of  the four  stars seem  to be
measurably  different  (Figure~\ref{fig:plx}).    Taking  the  average
parallax of C and D, 25.94\,mas,  as the best estimate of the distance
to the system (38.55\,pc), the parallax of A is lager by 1.28$\pm$0.58
mas.  This formally significant (2.2$\sigma$) difference translates to
the distance of  A 1.8$\pm$0.8 pc closer than C  and D.  However, Gaia
DR2 measured for A an  even more discrepant parallax of 28.94$\pm$0.58
mas.  The  inconsistency between  two Gaia  data releases  indicates a
problematic astrometry,  also corroborated by the  large RUWE.  During
the 2014.6-2017.4  period covered  by the EDR3,  the photocenter  of A
moved  almost linearly  in declination,  but its  motion in  RA had  a
substantial acceleration of 1.4  mas~yr$^{-2}$ ~according to the Aa,Ab
orbit.  Fitting the 5-parameter  solution (position, PM, and parallax)
to  this  non-linear  motion,  sampled   by  the  Gaia  scanning  law,
inevitably biases the parallax.  I tried to reproduce this effect, but
found a much  smaller ($<$0.1 mas) bias.  My toy  model also failed to
explain the difference between DR2 and EDR3 parallaxes of A.  The Gaia
data release  3 will  account for  accelerations and,  hopefully, will
give a more trustworthy parallax of A.  Cases where Gaia parallaxes of
stars in wide  physical binaries appear different because  one of them
contains an unresolved subsystem are not rare.

The   parallax   of  B   should   also   be   biased  by   the   Ba,Bb
subsystem. However, a smaller RUWE, a better agreement between the two
Gaia data releases (the DR2 parallax  of B is 25.94$\pm$0.14 mas), and
a smaller difference  from the mean parallax of C  and D indicate that
the problem is less severe, compared to star A.  The orbital period of
Ba,Bb estimated  from its projected separation  is on the order  of 10
yr, and the  actual period can be  longer or shorter by  2-3 times. If
the  period is  only  a few  years,  the impact  of  the subsystem  on
astrometry is  substantially reduced  by time averaging.  Although the
EDR3  parallax of  B differs  from the  mean parallax  of C  and D  by
0.36$\pm$0.09 mas,  I attribute this formally  significant discrepancy
to the bias.

%---------------------------------------------------------
\subsection{A String?}
\label{sec:string}

\begin{figure}
\epsscale{1.1}
\plotone{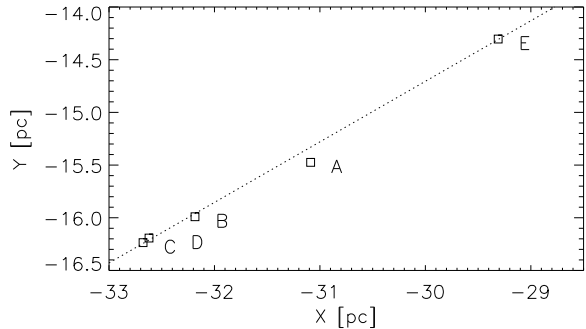}
%\plotone{XYplot.ps}
\caption{Location  of  stars A  to  E  projected onto  the Galactic  plane,
  assuming their EDR3 parallaxes. The dotted line is a linear fit.
\label{fig:XY}
}
\end{figure}

Suppose for the moment that the Gaia EDR3 parallaxes of A and B can be
trusted, implying different distances to these stars. Then their close
location  on  the   sky  is  a  mere  projection,   while  the  actual
configuration   in  space   is  a   line  pointing   toward  the   Sun
(Figure~\ref{fig:XY}).    In  such   case,   the   system  cannot   be
gravitationally bound, and there is no  reason to exclude from it star
E.  \citet{Kounkel2019} found  that young stars are  often arranged in
linear configurations,  strings.  However,  their strings  extend over
tens and hundreds of pc.

It appears  highly improbable  that four cluster  members accidentally
arranged themselves along  our line of sight. Even if  this were true,
there should be other members of this cluster around, but I found only
star E within a 5\degr ~search radius. Therefore, the V1311 Ori family
(excluding star  E) must be  a gravitationally bound  multiple system,
and its  configuration in  Figure~\ref{fig:XY} does not  correspond to
reality.

%---------------------------------------------------------
\subsection{Internal Motions}
\label{sec:internal}

At the  distance of 39\,pc,  the PM of  1 \masyr ~corresponds  to 0.18
\kms.   In principle,  accurate Gaia  astrometry can  measure relative
tangential motions in V1311 Ori  with a high precision, enabling study
of  its internal  kinematics. However,  the Gaia  PMs of  A and  B are
biased by inner subsystems.  The less  accurate long-term PM of star A
from Tycho-2 (Table~\ref{tab:1})  matches the PMs of  other stars, and
its difference with  the short-term Gaia PM  approximately matches the
orbital motion  of Aa,Ab (section~\ref{sec:orb}).  However,  the orbit
of  Ba,Bb is  not known  yet,  and its  PM  has not  been measured  by
Tycho-2.  UCAC4  gives the PM  of B as  (4.2$\pm$6.6, $-$34.8$\pm$4.9)
\masyr, while \citet{Vrijmoet2020} measured  for B (their component A)
a  PM of  (14.5$\pm$6.4, $-$50.5$\pm$2.9)  \masyr  ~on a  3.3 yr  time
base. The PM of B measured by Gaia DR2 and EDR3 is mutually consistent
to within 1 \masyr; it is  adopted here, despite an almost certain but
unknown  bias. In  contrast, the  EDR3 astrometry  of C  and D  can be
trusted. Their small  RUWE and matching RVs  speak against subsystems,
although cannot rule them out.

Neglecting the discrepant parallaxes of A  and B, I postulate that all
four stars are located  at a common distance of 38.55  pc. The PM bias
of  A and  B  caused  by their  subsystems  is  the largest  remaining
uncertainty  in  the study  of  the  internal kinematics;  another  is
related  to  the  estimated  masses. Despite  these  caveats,  we  can
evaluate whether this system can be bound or not.

Assume a binary  on a circular face-on  orbit with a period  $P$ and an
angular  separation (semimajor  axis)  $\rho$. Its  orbital speed  (in
arcsec~yr$^{-1}$) is
\begin{equation}
\mu^* =  (2 \pi \rho)/P = 2 \pi \rho^{-1/2} \varpi^{3/2} M^{1/2},
\label{eq:mu*}
\end{equation}
where $\varpi$ is  the parallax in arcseconds, $M$ is  the mass sum in
solar   units,   and  the   third   Keler's   law   is  used,   $P   =
(\rho/\varpi)^{3/2} M^{-1/2}$.   The characteristic speed $\mu^*$  is a
scaling  factor for  binaries  with arbitrary  eccentricity and  orbit
orientation.   In  a bound  binary  with  negative total  energy,  the
relative speed $\Delta \mu$ is always less than $\sqrt{2} \mu^*$. This
is a necessary (but not sufficient) condition of boundness.

\begin{table}
\center
\caption{Internal kinematics of V1311 Ori}
\label{tab:kin}
%\medskip
\begin{tabular}{l c  c  c  cc} 
\hline
Pair   & $\rho$ & $\theta$ &  $s$   & $\Delta \mu$ & $\mu^*$ \\   
       &(arcsec) & (deg)   & (au) &   \multicolumn{2}{c}{ (mas yr$^{-1}$)}        \\
\hline
B,C      & 5.31   & 125.5  & 205  &  9.73 (4.53)   & 9.72  \\
A,BC     & 148.5  & 318.6  & 5724 &  8.15 (1.81)   & 2.91   \\
ABC,D    & 217.5  & 16.3   & 8385  & 2.00 (0.86)   & 2.51 \\
A,D      & 254.2  & 4.9    & 9799 &  1.53          & 1.86 \\  
C,D      & 185.5  & 39.1   & 9799 &  3.36          & 1.41 \\     
\hline
\end{tabular}
\end{table}

Adopting the masses  of 1.1, 0.50, 0.23, and 0.17  \msun ~for stars A,
B, C, and D, I computed the mass-weighted positions and PMs of various
combinations,   their   relative   motion  $\Delta   \mu$,   and   the
characteristic  speed $\mu^*$.   Representative results  are given  in
Table~\ref{tab:kin}. In  brackets, $\Delta \mu$ is  computed using the
UCAC4  $\mu_\alpha^* =  4.2$ \masyr  ~for  B instead  of 17.62  \masyr
~measured by  Gaia EDR3.   The 5\farcs3  pair B,C  is likely  close in
space  (not  only  in  projection).   Its  relative  motion  does  not
contradict the  bound status regardless of  the adopted $\mu_\alpha^*$
of star  B.  However, a similar  test applied the A,BC  pair indicates
that it can be  bound with the UCAC4 PM of B, but  is unbound with the
EDR3 PM.  The widest combination ABC,D looks bound in both cases.  The
two lowest-mass  stars C and  D with  accurate astrometry cannot  be a
bound pair,  which is natural  (C moves  too fast because  it revolves
around B). On the other hand, A,D  could be a wide bound pair (in fact
triple) that projects on to another bound triple B,C.

The  hierarchical structure  shown in  Figure~\ref{fig:mobile} assumes
that D is  the outer component in this sextuple  system.  Its distance
from ABC along the line of sight must be at least 15 kau to ensure the
dynamical stability (the  accurate parallaxes of C and  D indeed imply
such distance difference, with  a low significance).  Even considering
the line-of-sight distance  of D, the tangential motion  of D relative
to ABC is slow enough for  a bound system.  Alternatively, this system
can  be  dynamically unstable.   With  an  estimated outer  period  of
$\sim$1 Myr,  it could  have survived for  several crossing  times and
might disrupt in the future.

%---------------------------------------------------------
\subsection{Galactic Motion}
\label{sec:gal}

\begin{deluxetable}{l ccc} 
\tablecaption{Galactic motion
\label{tab:gal} }
\tablewidth{0pt}                                   
\tablehead{   
\colhead{Star} &
\colhead{$U$} &
\colhead{$V$} &
\colhead{$W$} \\
 & 
\multicolumn{3}{c}{\kms}
}
\startdata
A    & $-$14.1 &  $-$16.4 &  $-$10.2 \\
B    & $-$14.8 &  $-$18.2 &   $-$9.0 \\
C    & $-$14.5 &  $-$17.7 &  $-$10.7 \\
D    & $-$15.2 &  $-$17.4 &  $-$10.5 \\
E    & $-$13.1 &  $-$16.6 &   $-$9.5 \\
BPMG & $-$10.9 &  $-$16.0  & $-$9.0 \\
THOR & $-$12.8 &  $-$17.7  & $-$9.0 
\enddata
\end{deluxetable}

Using data on individual stars from Table~\ref{tab:1} (including E), I
compute the  Galactic velocities $U,V,W$  (the $U$ axis  points toward
the Galactic center)  and list the result  in Table~\ref{tab:gal}. The
unknown RV of C  is assumed to equal the mean RV  of the system.  Mean
velocities of the BPMG and THOR moving groups are given for reference,
according to  \citet{Gagne2018}. The  internal velocity  dispersion in
these groups  is about 1 \kms,  and their mean velocities  also differ
between authors by similar amounts.   V1311 Ori is located at Galactic
coordinates   $(l,b)  =   (206\fdg5,-19\fdg0)$,   roughly  in   the
anti-center  direction,  so  the  RV  errors  affect  mostly  the  $U$
component.  The  velocities of the four  stars, computed independently
of each  other, are  mutually consistent  and closer  to THOR  than to
BPMG.  Given  the similar kinematics  and age  of both YMGs,  they are
likely related to  a common star formation region, to  which V1311 Ori
also  belonged   \citep{Gagne2021}.   Both  the  kinematics   and  the
isochrones confirm the age of $\sim$24 Myr for the V1311 Ori group.

Motion  with  a  relative  velocity  of  1  \kms,  typical  for  YMGs,
corresponds to  1 pc~Myr$^{-1}$,  so stars  born together  have little
chance to stay in a volume of 0.05  pc radius during 24 Myr. This is a
strong argument favoring the bound nature of the V1311 Ori system.  In
contrast, star  E (RX J0534.0-0221) is  an unbound member of  the YMG,
separated from V1311 Ori  along the line of sight by  4\,pc. The RV of
star E differs  from the RVs of  other stars by 1-2 \kms,  so it could
cover this distance in 2-4 Myr.  However, in the tangential plane E is
moving towards V1311 Ori, making it highly unlikely that E was ejected
from this system a few Myr ago.

%---------------------------------------------------------
\section{Discussion and Summary}
\label{sec:sum}

The family of  V1311 Ori is a gravitationally  bound system containing
six low-mass PMS stars. Its spatial  motion is similar to the BPMG and
THOR  moving   groups,  so   V1311  Ori   originated  from   the  same
star-formation region  some 24 Myr  ago. Fast rotation of  some stars,
their location on the CMD,  chromopsheric and X-ray emission match the
young age. 

Although the  Gaia parallaxes and PMs  of the brightest stars  A and B
are  biased   by  subsystems,  the  internal   motions  inferred  from
astrometry appear to be slow and do not contradict the bound nature of
this  system of  $\sim$10 kau  size.  Its  structure, however,  is not
clear.  The projected separations between  A, BC, and D are comparable
(Figure~\ref{fig:sky}),  so   it  could  be  a   dynamically  unstable
configuration, in other words  a mini-cluster.  The estimated crossing
time on the order of 1 Myr suggests that the system has survived until
now, but  may evolve  in the future  owing to  dynamical interactions
between its members.   One or both lowest-mass stars C  and D might be
ejected, leaving  a stable hierarchy  with 4-5 components.   The tight
inner pairs Aa,Ab and Ba,Bb will not be affected.  Alternatively, this
system could be already dynamically stable, with a hierarchy described
by Figure~\ref{fig:mobile}, if star D is  closer by $>$15 kau than ABC.
Other configurations  (for example,  two triples Aa,Ab-D  and Ba,Bb-C)
are not excluded, but appear less likely.

The V1311 Ori system was discovered in the search for wide hierarchies
within 100 pc \citep{triads}.  Some low-mass wide triples in the field
also have non-hierarchical configurations with comparable separations.
Admittedly,  a  stable  triple  can  appear  non-hierarchical  due  to
projection, but statistical analysis of all these systems demonstrates
that many  are indeed  just above the  stability limit.   They witness
early dynamical interactions in unstable hierarchies and represent the
surviving population.  The system V1311 Ori illustrates the transition
from assembly to dispersal.

\citet{Gagne2021} establish the relation  between BPMG and THOR groups
and believe that they were formed  together, as well as the Kounkel \&
Covey's  groups Theia  62 and  65.  The  large size  of the  V1311 Ori
system  ($\sim$10  kau)  speaks  against  its  formation  in  a  dense
environment.   Small masses  of  these M-dwarfs  do  not favor  binary
formation   by  disk   fragmentation  \citep{Kratter2016}.    So,  the
architecture of  this and other similar  low-mass hierarchies reflects
only  three  basic processes  involved  in  the formation  of  stellar
systems: fragmentation,  accretion, and internal dynamics,  while disk
fragmentation and  dynamical interactions  with other  cluster members
are irrelevant.  Fragmentation and collapse begin in the densest parts
of the parent cloud \citep{Vasquez2019},  and these first stars have a
larger supply of gas, compared to stars formed later at the periphery.
An accreting binary shrinks while its mass ratio increases.  The inner
and most  massive subsystems Aa,Ab and  Ba,Bb in V1311 Ori  with large
mass  ratios were  likely the  first to  form. Stars  C and  D, formed
later, have smaller masses.  They  could be gravitationally bound to A
and B from the outset if the internal motions in the parent cloud were
slow, or  became bound  as they  approached and  got captured  on wide
orbits, possibly with assistance of the  remaining gas around A and B.
As a result, the system of  V1311 Ori is mass-segregated, resembling a
young custer. In fact, it is (or was) a cluster with a small number of
stars.  Other  wide low-mass  marginally stable  triples in  the field
could  be the  remnants of  similar mini-clusters.   They also  appear
mass-segregared (in two thirds of those triples, the most massive star
belongs to the inner pair).

This work  is devoted to the  structure and dynamics of  the V1311 Ori
system.   The physics  of these  young low-mass  stars is  outside its
scope, but  it is definitely worth  further study. Being members  of a
coeval group with a well-measured  distance, they are more interesting
than  simple stars  or binaries.   Measuring stellar  masses from  the
orbits of Aa,Ab and Ba,Bb is an obvious prospect.

%\acknowledgements
\begin{acknowledgments} 

The research  was funded  by the  NSF's NOIRLab.   This work  used the
SIMBAD   service  operated   by   Centre   des  Donn\'ees   Stellaires
(Strasbourg, France),  bibliographic references from  the Astrophysics
Data System  maintained by SAO/NASA.  This  work has made use  of data
from    the    European    Space    Agency    (ESA)    mission    Gaia
(\url{https://www.cosmos.esa.int/gaia}),  processed by  the Gaia  Data
Processing        and         Analysis        Consortium        (DPAC,
\url{https://www.cosmos.esa.int/web/gaia/dpac/consortium}).    Funding
for the DPAC has been provided by national institutions, in particular
the  institutions participating  in the  Gaia Multilateral  Agreement.
This paper includes  data collected by the TESS mission  funded by the
NASA Explorer  Program.  The  data were  obtained through  the Barbara
A. Mikulski Archive for Space  Telescopes  (MAST) at the   
 Space Telescope Science Institute. The specific observations analyzed
 can be accessed via
 \dataset[10.17909/t9-nmc8-f686]{https://doi.org/10.17909/t9-nmc8-f686}. 
This research has made use
of the  services of the  ESO Science Archive Facility  (FEROS spectra,
ESO programs 094.A-9002 and 089.A-913).

\end{acknowledgments} 

\facility{SOAR, CTIO:1.5m, Gaia, TESS}

%---------------------------------------------------------
%\section{}
%\label{sec:}

%---------------------------------------------------------
%\section{}
%\label{sec:}

%---------------------------------------------------------
%\section{}
%\label{sec:}

%\subsection{}

%\bibliography{v1311ori.bib}
%\bibliographystyle{aasjournal}
%\bibliographystyle{apj}

\end{document}